\documentclass[aps,prl,reprint,superscriptaddress]{revtex4-2}

\usepackage[T1]{fontenc}
\usepackage[utf8]{inputenc}
\usepackage{lmodern}
\usepackage{amsmath,amssymb,bm}
\usepackage{graphicx}
\usepackage[colorlinks=true,citecolor=blue,linkcolor=blue,urlcolor=blue]{hyperref}

\begin{document}

\title{Wave Resistance for Stochastic Motion at Interfaces}

\author{Maxence Arutkin}
\affiliation{School of Chemistry, Center for the Physics \& Chemistry of Living Systems, Sackler Center for Computational Molecular \& Materials Science, Tel Aviv University, 6997801, Tel Aviv, Israel}
\author{Shlomi Reuveni}
\affiliation{School of Chemistry, Center for the Physics \& Chemistry of Living Systems, Sackler Center for Computational Molecular \& Materials Science, Tel Aviv University, 6997801, Tel Aviv, Israel}
\author{Elie Rapha\"el}
\affiliation{Gulliver UMR CNRS 7083, ESPCI Paris, PSL Research University, 75005 Paris, France}

\date{\today}

\begin{abstract}
Wave resistance is the drag generated by the wave radiation that a source moving at a fluid interface sustains. Under stochastic trajectories, the mean drag is controlled by the ensemble-averaged surface profile built from the trajectory history. We show that the result is a finite resistance below the deterministic radiation threshold and a regularization of the singular response at the minimum phase velocity of the capillary-gravity waves. We derive explicit scaling laws for drifted Brownian trajectories, including a universal high-diffusivity decay. For drifted L\'evy flight, we find the mean wave resistance in closed-form, extending wave-drag theory to non-Gaussian trajectories. 
\end{abstract}

\maketitle

Wave resistance, the drag a moving disturbance experiences at a fluid interface, is a canonical problem in interfacial hydrodynamics. When a localized pressure distribution moves steadily along the air–water interface, it radiates capillary–gravity waves that carry energy and momentum to infinity. Havelock calculated this force exactly for uniform steady motion~\cite{havelock1932theory}. This result has been subsequently applied to systems ranging from ship hulls to walking insects~\cite{raphael1996capillary,chepelianskii2010self,hu2003hydrodynamics,steinmann2018unsteady,voise2010management,kim2024,hunt2023drag,jami2021overcoming}. However, many interfacial phenomena such as Brownian colloids undergoing thermal fluctuations~\cite{villa2023brownian,villa2023prolate}, active swimmers undergoing stochastic reorientation~\cite{tailleur2008statistical,cates2015motility,willems2025,dhar2024,nambiar2024,deng2022}, or whirligig beetles performing erratic search patterns~\cite{devereux2021whirligig}, involve stochastic trajectories for which no ensemble-level wave-drag theory exists.

The difficulty is that wave resistance depends on the entire trajectory history: waves radiated at time $t-\tau$ continue to propagate and interfere with the current source position, creating history-dependent hydrodynamic interactions mediated by the dispersive wave field. For steady motion at velocity $v$, this history-dependence simplifies: if $v<c_{\min}=(4g\gamma/\rho)^{1/4}$, with $g, \gamma$ and $\rho$ standing for the gravitational acceleration, surface tension coefficient and fluid density, respectively, ($c_{\min}\approx 0.23$~m/s for water at room temperature) no resonant wavenumber exists and the resistance vanishes.  For $v>c_{\min}$ a stationary wave pattern forms and Havelock's formula applies~\cite{raphael1996capillary}. However, when the position of the source is a random process governed by diffusion, reorientation, or active forcing, the coupling between velocity fluctuations and dispersive propagation becomes nontrivial, reshaping the mean wave field around the source, and rendering the resulting drag inaccessible from deterministic solutions. Classical singularities at the threshold velocity, where phase and group velocities coincide~\cite{raphael1996capillary,benzaquen2011wave}, further complicate the analysis.

The framework of Gierczak \textit{et al.}~\cite{gierczak2020unsteady} provides the foundations for 2D time-dependent wave resistance~\cite{dode2022wave}. Yet, ensemble averaging over stochastic trajectories remains an open problem. In this letter, we reduce the retarded force integral to one dimension, which permits exact analytical treatment and explicit ensemble averaging over arbitrary stochastic processes. This extension enables us to quantify how velocity fluctuations couple to the dispersive wave field and to reveal asymptotic regimes of stochastic wave resistance.

We first consider Brownian motion with drift, for which the ensemble-averaged resistive force, $\langle R \rangle$ in Eq.~\ref{eq:Rmean}, exhibits a rich phase diagram in the velocity–diffusivity plane. For $v>c_{\min}$, the Brownian kernel recovers Havelock's resistance in the deterministic limit $\Delta\to0$ (with $\Delta$ being a dimensionless diffusion coefficient defined below), the detailed supercritical small-$\Delta$ regime approaches the Havelock limit and is not developed here. For $v < c_{\min}$, no Havelock resonance exists; the drag instead arises from diffusive symmetry breaking of the mean surface deformation, and, close to the threshold, from a stationary-point contribution rescued by Brownian decorrelation. These competing effects generate three asymptotic regimes, including a universal scaling $\langle R \rangle \sim v\,\Delta^{-5/3}$ at high diffusivity (Eq.~\ref{eq:regI}) and a threshold regularization $\langle R \rangle \sim \Delta^{-1/2}$ at $v=c_{\min}$ (Eq.~\ref{eq:threshold}). 

\textit{Generalized wave resistance---}We begin by deriving the general resistance formula for arbitrary one-dimensional trajectories, specializing the two-dimensional framework of Gierczak \textit{et al.}~\cite{gierczak2020unsteady} to the line-load geometry of Rapha\"el \& de Gennes~\cite{raphael1996capillary}. This one-dimensional reduction enables analytical treatment while retaining the essential physics of dispersive wave coupling.

We consider a localized pressure distribution $P(x,t)=P_0(x-x_t)$ moving along an arbitrary trajectory $x_t$ at a deep-water interface supporting capillary-gravity waves with dispersion relation $\omega^2(k)=gk+\gamma k^3/\rho$. The wave resistance is defined as the force required to maintain the prescribed motion of the source and is computed as the pressure acting against the slope of the surface~\cite{havelock1932theory}, $R(t)=\int P(x,t)\partial_x\eta(x,t)\,dx$, where  $\eta(x,t)$ is the surface displacement. Assuming incompressible, inviscid flow and small-amplitude deformations, linearized free-surface boundary conditions (kinematic and dynamic) yield the horizontal reaction force $R(t)$ on the moving disturbance. The solution takes the form of a retarded integral over the trajectory history (Supplemental Material \footnote{See Supplemental Material for derivations of the retarded force integral, Brownian regime asymptotics, and mean-profile formula.}):
\begin{equation}
R(t)=\frac{1}{\pi\rho}\int_{0}^{t}\!d\tau\int_{0}^{\infty}\!dk\,f(k)
\frac{\sin(\omega(k)\tau)}{\omega(k)}\,
\sin(k(x_t-x_{t-\tau})),
\label{eq:kernel}
\end{equation}
with $f(k)=k^2|\hat P(k)|^2$, where $\hat P(k)$ is the Fourier spectrum of the arbitrary pressure profile. This is the 1D version of the generalized formula derived in \cite{gierczak2020unsteady}. 

For concrete analysis, we model the disturbance as a Lorentzian pressure field, whose Fourier spectrum satisfies $\hat P(k) \propto e^{-a\vert k\vert}$ (the exponential decay in Fourier space acts as a smooth large-$k$ cutoff). Replacing it by a Gaussian pressure field leaves the scaling exponents of the asymptotic regimes unchanged. The sine factor $\sin(k(x_t-x_{t-\tau}))$ encodes the phase accumulated by each wavenumber $k$ when going from $t-\tau$ to $t$. This accumulated phase determines whether waves emitted previously interfere constructively or destructively with the present motion of the source.

For steady translation $x_t=vt$, the trajectory increment becomes $x_t-x_{t-\tau}=v\tau$, and the stationary-phase condition selects resonant wavenumbers satisfying $\omega(k)=kv$. In this limit, the $\tau$-integral yields $\frac{\pi}{2\omega(k)}\delta(\omega(k)-kv)$ as $t\to\infty$, exactly recovering Havelock’s formula for the steady drag $R_{\mathrm{steady}}(v) = \frac{1}{2\rho}\int_{0}^{\infty}\!dk\,
\frac{k^2\,\big|\widehat P(k)\big|^2}{\omega(k)}\,\delta \left(\omega(k)-kv\right)$~\cite{havelock1932theory}. More generally, 
Eq.~\eqref{eq:kernel} provides the general instantaneous resistance for any source trajectory $x_t$, without restriction on velocity profile, acceleration, or statistical properties.

We now turn to apply this general framework to stochastic trajectories. The kernel in Eq.~\eqref{eq:kernel} depends on the path only through the increments $\Delta x(\tau)=x_t-x_{t-\tau}$. For processes with stationary increments, ensemble averaging therefore reduces to evaluating the increment characteristic function $\varphi(k,\tau)=\big\langle e^{ik\Delta x(\tau)}\big\rangle$ (equivalently $\big\langle \sin(k\Delta x(\tau))\big\rangle=\mathrm{Im}\,\varphi(k,\tau)$), after which the $\tau$-integral yields a stationary mean resistance in the long-time limit. This provides a direct route from a stochastic kinematics model (i.e. $\varphi$) to an explicit wave-drag law. We first illustrate this on drifted Brownian motion, where $\varphi$ is known in closed form and the mean wave resistance can be derived analytically.

\textit{Brownian averaging and mean resistance---}Brownian motion with drift, where $v$ is imposed by the source's own motion and not by a background flow, combines diffusion with directed transport, providing the natural first application. Its Gaussian statistics permit exact ensemble averaging and asymptotic analysis, revealing how stochasticity modifies the deterministic Havelock result.

We consider $x(t)=vt+B_t$, where $B_t$ is the Brownian motion with $\langle(B_t-B_{t-\tau})^{2}\rangle=2D\tau$, and $\langle\cdot\rangle$ denoting the ensemble average. Averaging in the comoving frame yields $\langle R(t)\rangle = \int P_0(x)\langle \partial_x\eta(x,t)\rangle\,\mathrm{d}x$, the mean wave resistance is the integral of the pressure field against the slope of the mean profile. Since trajectory increments are Gaussian, the ensemble average factorizes to
\begin{equation}
\big\langle \sin(k(x_t-x_{t-\tau}))\big\rangle
=\sin(kv\tau)\,e^{-Dk^{2}\tau}.
\label{eq:gauss}
\end{equation}
The resistance $R(t)$ depends only on trajectory increments $x_t-x_{t-\tau}$, whose statistics are time-independent, yielding a statistically stationary observable in the limit $t\to\infty$. Substituting Eq.~\eqref{eq:gauss} into Eq.~\eqref{eq:kernel} and performing the time integral over $\tau$ yields the mean resistance
\begin{equation}
\langle R\rangle=\frac{1}{\pi\rho}\int_{0}^{\infty}\!dk\,k^{2}|\hat P(k)|^{2}\,\mathcal{K}(k;v,D),
\label{eq:Rmean}
\end{equation}
where the integrand $\mathcal{K}$ encodes velocity–diffusion coupling
\begin{equation}
\mathcal{K}(k;v,D)=
\frac{2Dk^{3}v}{
[(Dk^{2})^{2}+(\omega-vk)^{2}]
[(Dk^{2})^{2}+(\omega+vk)^{2}]}.
\label{eq:kernel_BM_correct}
\end{equation}

The denominator reveals Lorentzian broadening of the classical Havelock resonance $\omega(k)=kv$: diffusion regularizes the $v=c_{\min}$ singularity by introducing a finite correlation time. The factor $(\omega-vk)$ vanishes on the Havelock resonance $\omega(k)=vk$ and drives the resonant response, while $(\omega+vk)$ stays strictly positive for all physical wavenumbers and contributes only a smooth non-resonant background.

In what follows, we nondimensionalize velocities by $c_{\min}=(4g\gamma/\rho)^{1/4}$, wavenumbers by $k_c=\sqrt{\rho g/\gamma}$, and diffusivity by $\Delta=Dk_c/c_{\min}$, yielding a dimensionless resistance $\mathcal{R}(v,\Delta)$, and dimensionless dispersion relation $\omega^2(k) = \frac{1}{2}(k+k^3)$ (Supplemental Material). The resulting behavior is governed by the competition between the diffusive decorrelation time $\tau_D\sim(Dk^{2})^{-1}$ and the wave period $\tau_{\mathrm{wave}}\sim\omega^{-1}(k)$.

\begin{figure}[t] 
    \centering 
    \includegraphics[width=0.9\linewidth]{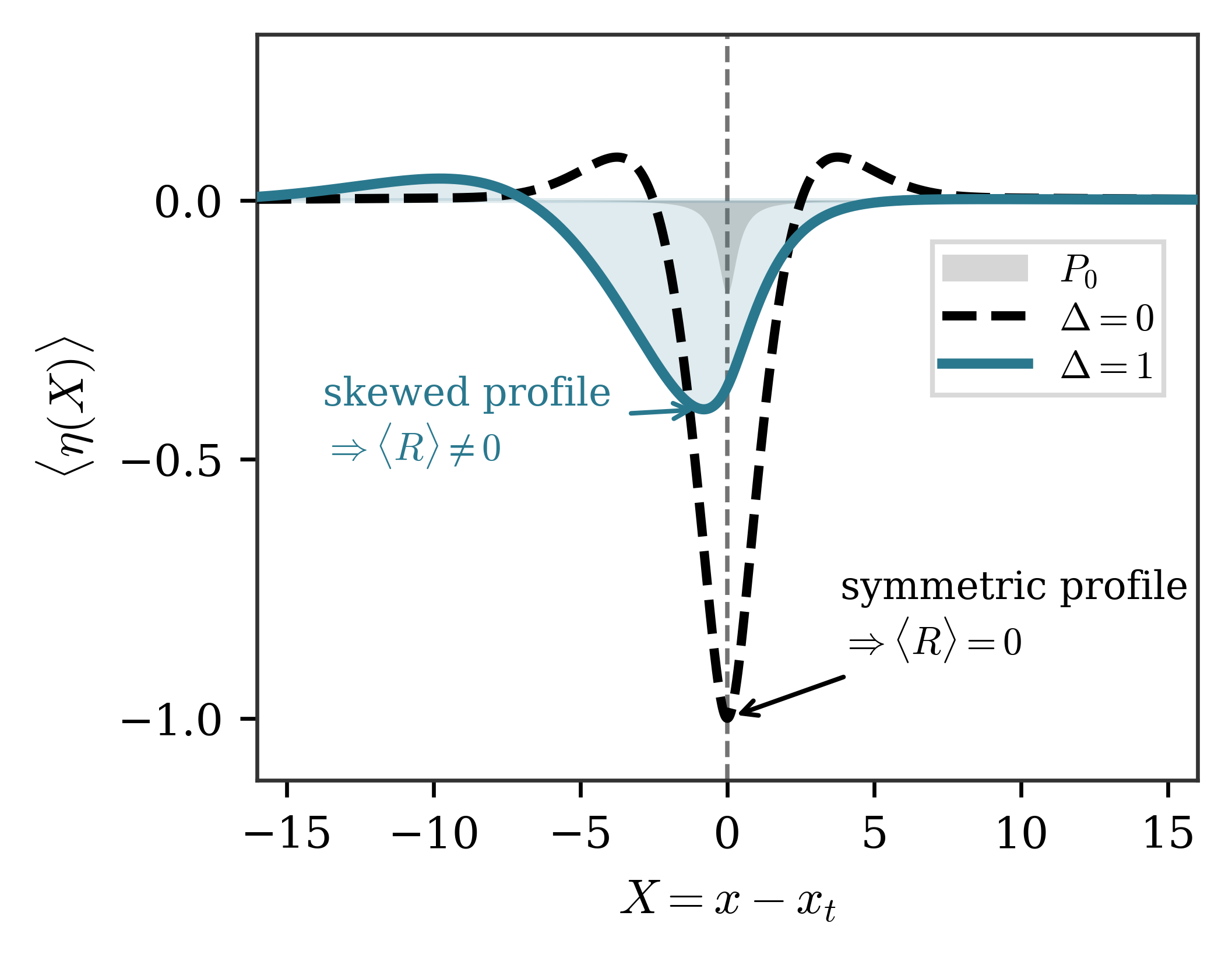} 
    \caption{\textbf{Mean-profile asymmetry.} Mean surface $\langle\eta(X)\rangle$ in the comoving frame at $v=0.8$ for $\Delta=0$ (deterministic, symmetric, no drag) and $\Delta=1$ (diffusive, skewed, finite drag). Pressure $P_0$ shaded in gray. Lorentzian source, $a=0.20$.} \label{fig:profile_mechanism}
\end{figure}

\textit{Asymmetry of the mean profile---}A nonzero mean wave resistance requires an asymmetric mean profile. Figure~\ref{fig:profile_mechanism} shows the case $v=0.8$. At $\Delta=0$ the profile is symmetric and the drag is zero, at finite $\Delta$ the profile becomes asymmetric and the drag is nonzero. In Fig. S1, we provide profiles for $v\geq1$, to illustrate how diffusion modifies the resulting wave pattern. In the absence of diffusion, the profile has capillary waves ahead and gravity waves behind the source. Finite $\Delta$ damps these trains, especially the shorter-wavelength capillary part. For $\Delta \gg 1$, the profiles no longer show capillary or gravity wave trains. Averaging leaves a skewed deformation around the source, the skew is set by the drift direction and vanishes at $v=0$.

\begin{figure}[t]
  \centering
  \includegraphics[width=0.9\linewidth]{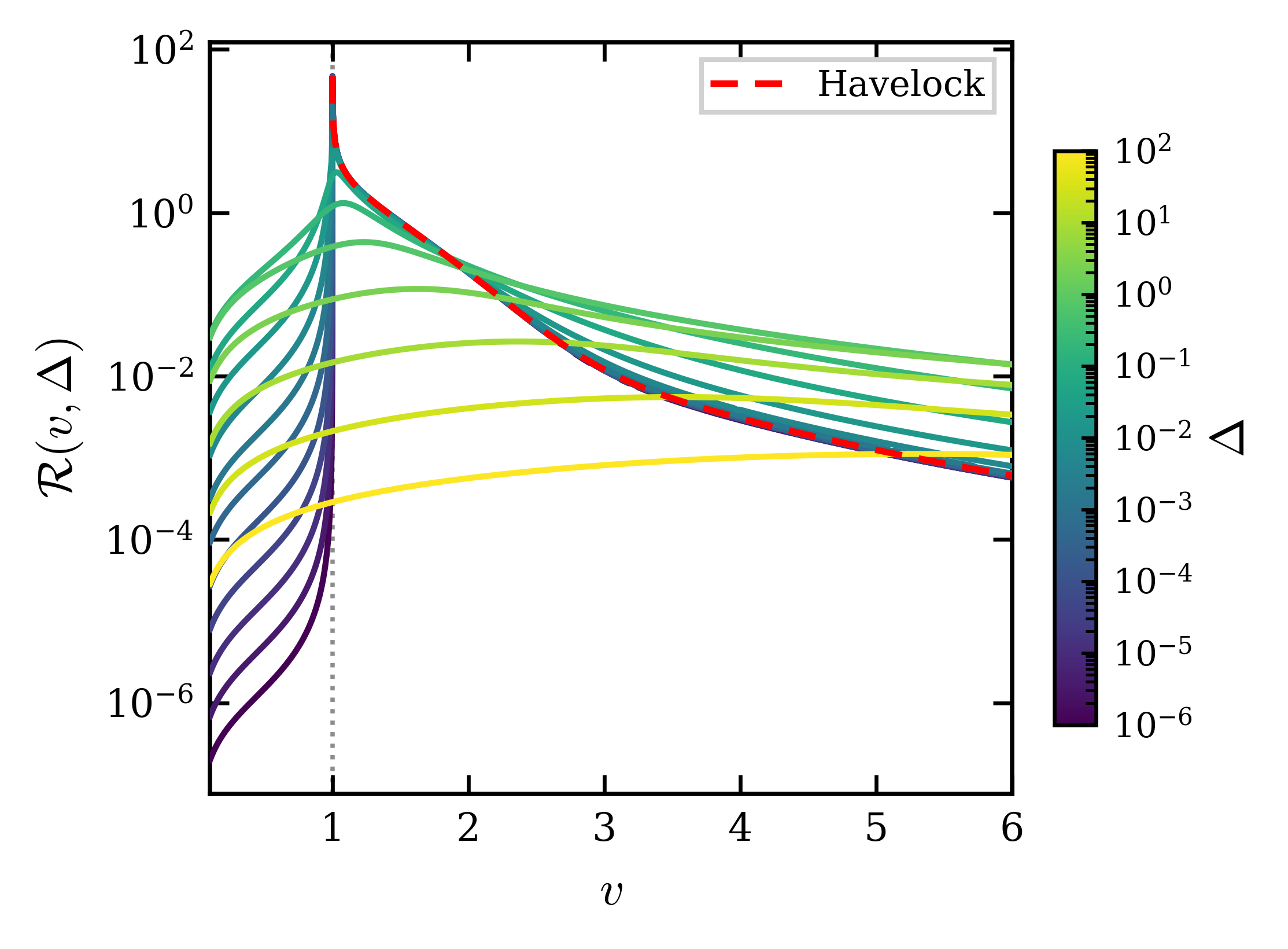}
  \caption{\textbf{Mean wave resistance versus drift speed across diffusivities.} Wave resistance $\mathcal{R}(v, \Delta)$ versus dimensionless drift speed $v$ for Brownian motion with Lorentzian pressure field ($a = 0.20$). Solid curves: exact evaluation of Eq.~\eqref{eq:Rmean} for sixteen diffusivities $\Delta \in [10^{-6}, 10^{2}]$. Red dashed line: classical Havelock resistance ($v > 1$), recovered as $\Delta \to 0$. The corresponding mean surface profiles are displayed in the Fig.~S1 in the Supplemental Material.}
  \label{fig:R_vs_v}
\end{figure}

\begin{figure}[t]
  \centering
  \includegraphics[width=0.9\linewidth]{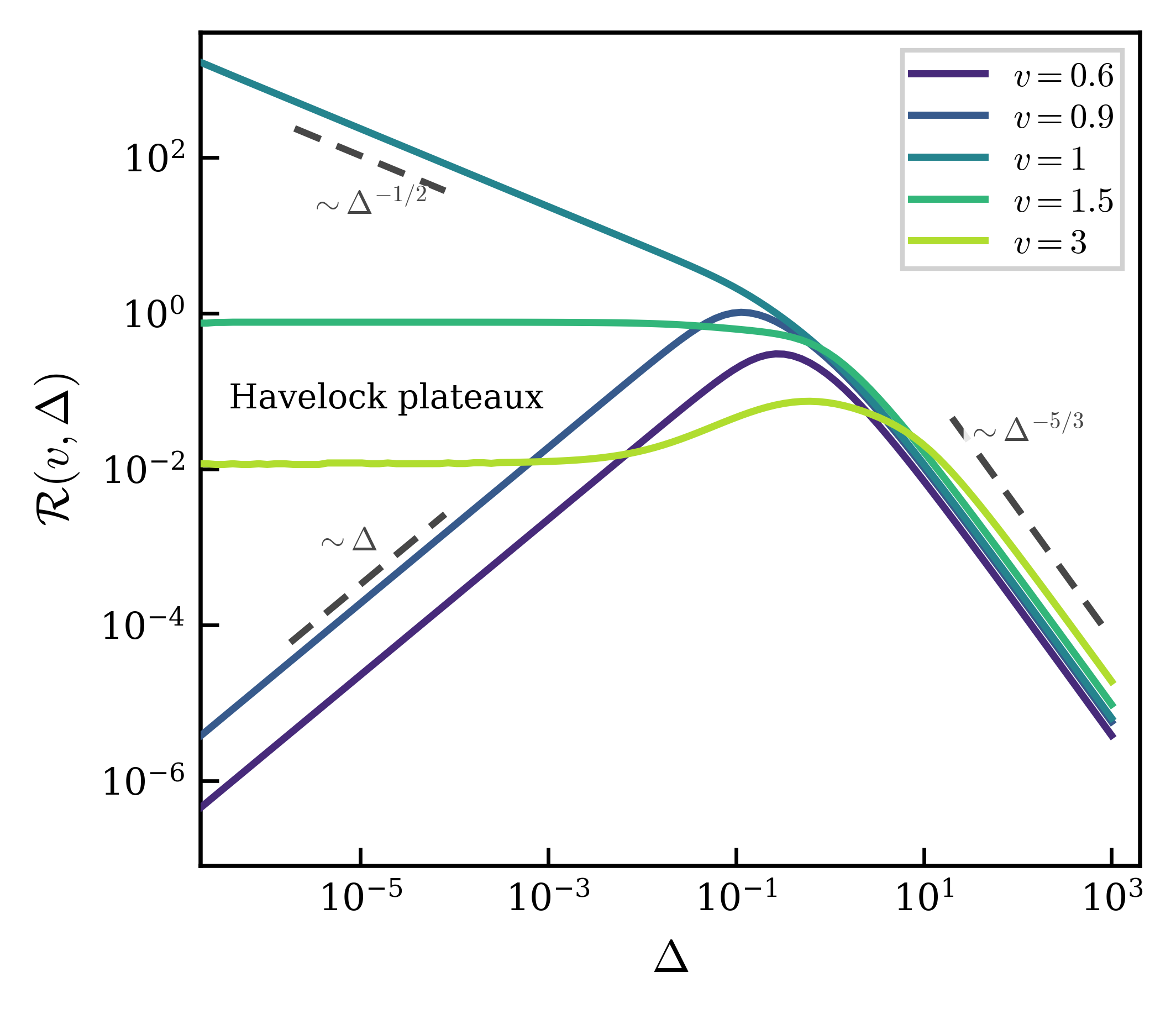}
  \caption{\textbf{Asymptotic scaling laws across diffusivity regimes.}
  Mean wave resistance $\mathcal{R}(v, \Delta)$ versus diffusivity $\Delta$ for five drift speeds $v \in [0.6, 3.0]$ ($a = 0.20$, log-log axes). Solid curves: exact evaluation of Eq.~\eqref{eq:Rmean}. At high diffusivities, all curves align sharing the universal $\sim\Delta^{-5/3}$ asymptotics of Eq.~\eqref{eq:regI} (Regime~I).  At low diffusivities, and at subcritical velocities, one obtains the $\sim\Delta$  asymptotics of Eq.~\eqref{eq:regII} (Regime~II), while at critical speed $v=1$, one obtains the $\sim\Delta^{-1/2}$  asymptotics of Eq.~\eqref{eq:threshold} evaluated at $\delta=0$. At low  diffusivities, and supercritical velocities, one approaches the Havelock limit which is insensitive to $\Delta$.}
  \label{fig:R_vs_Delta}
\end{figure}

\textit{Drifted Brownian asymptotic regimes---}Three asymptotic limits are developed below (also see Figs.~\ref{fig:R_vs_v}--\ref{fig:R_vs_Delta}): universal $\sim\Delta^{-5/3}$ scaling at high diffusivity (Regime I), linear growth at subcritical speeds (Regime II) and threshold regularization with $\mathcal{R}\sim\Delta^{-1/2}$ together with the subcritical threshold law (Regime III). The supercritical small-$\Delta$ regime reduces to the classical Havelock limit and is not expanded here.

\emph{Regime I (high diffusivity).}~When the dimensionless diffusivity $\Delta \gg 1$ at fixed drift speed $v$, the mean wave resistance exhibits universal scaling behavior independent of source geometry. The dominant contribution arises from small wavenumbers $k \ll 1$, where the dispersion relation simplifies to $\omega^2 \approx k/2$ and the pressure spectrum becomes approximately constant, $|\hat P(k)|^2 \approx |\hat P(0)|^2$. In this long-wavelength regime, the Lorentzian denominators in $\mathcal{K}$ contain two competing terms: $\Delta^2 k^4$ from diffusive decorrelation and $\omega^2(k) \sim k$ from wave propagation.

Asymptotic analysis shows that the integral is governed by the dominant balance $\Delta^2 k^4 \sim \omega^2(k) \sim k$, which selects the characteristic scale $k_\star \sim \Delta^{-2/3}$. Introducing the rescaled variable $y = k\Delta^{2/3}$ absorbs all $\Delta$-dependence into an overall prefactor (Supplemental Material), yielding
\begin{equation}
\mathcal{R}(v,\Delta)\sim C_\star\,v\,\Delta^{-5/3},
\label{eq:regI}
\end{equation}
where the universal constant $C_\star=\tfrac{2^{5/3}\pi}{9\sqrt{3}}$ arises from evaluating the rescaled integral over $y$. The $-5/3$ scaling characterizes the universal high-diffusivity limit.

\emph{Regime II (subcritical speeds, weak diffusion).} For subcritical drift speeds $v \ll 1$ in the ballistic limit $\Delta \to 0$, the resistance vanishes: no wavenumber satisfies the resonance condition $\omega(k) = vk$, and thus no steady waves are emitted. This holds for constant-velocity motion, i.e. vanishing acceleration at all times. As soon as acceleration exists (e.g. circular motion, sudden start) it can radiate below $c_{\min}$ \cite{chepelianskii2008capillary,closa2010capillary}. 

Asymptotic analysis (Supplemental Material) shows that the resistance grows linearly with drift and diffusivity,
\begin{equation}
\mathcal{R}(v,\Delta) \propto v\,\Delta.
\label{eq:regII}
\end{equation}
The detailed prefactor depends on the pressure-width parameter $a$ (see Table~S.II in the Supplemental Material): narrow sources ($a \ll \Delta^2$) exhibit logarithmic sensitivity to $\Delta$, whereas very broad sources ($a \gg \Delta^2$) produce an algebraic decay $\mathcal{R} \sim v \Delta a^{-4}$. In Fig.~\ref{fig:R_vs_Delta}, this linear-growth regime appears for subcritical curves at small $\Delta$, before crossing over to the universal $\Delta^{-5/3}$ tail at large diffusivity.

\emph{Regime III (threshold and subcritical threshold law).} The linear-growth regime at subcritical speeds terminates at the threshold $v = 1$, the minimum phase velocity $c_{\min} = (4g\gamma/\rho)^{1/4}$, where the deterministic resistance diverges. This minimum is a consequence of competing restoring forces, gravity and surface tension, whose relative importance reverses at $k_c = \sqrt{\rho g/\gamma}$. At this threshold the phase and group velocities coincide, $c_p(1) = c_g(1) = 1$, such that $\omega(k)-vk$ and its derivative with respect to $k$ vanish simultaneously at $k = 1$. For $v = 1 + \delta$ with $\delta \le 0$ and $|\delta| \ll 1$, stationary-phase analysis of the kernel near $k=1$ gives a closed-form expression for $\mathcal R(v,\Delta)$, valid at the threshold and on the subcritical, $\delta<0$, side (Supplemental Material)
\begin{equation}
\mathcal{R}(1+\delta,\Delta)\sim \frac{\pi}{4}\,e^{-2a}\,
\frac{\sqrt{S+\Delta} -\sqrt{S-\Delta}}{S},
\label{eq:threshold}
\end{equation}
with $S = \sqrt{\Delta^{2}+\delta^{2}}$. This expression organizes two distinct asymptotic limits. First, exactly at the threshold ($\delta = 0$), it reduces to $\mathcal{R}(1,\Delta) \sim (\pi/2\sqrt{2})e^{-2a}\Delta^{-1/2}$, where the universal exponent $-1/2$ emerges from the quadratic expansion of the dispersion relation near $k=1$. Second, on the subcritical side ($\delta < 0$, $|\delta| \gg \Delta$), it follows $\mathcal{R} \sim \Delta|\delta|^{-3/2}$. With $x=\delta/\Delta$ and $Y=(2/\pi)\,e^{2a}\,\sqrt{S}\,\mathcal{R}$, Eq.~\eqref{eq:threshold} gives $Y=\cos \left(\frac{1}{2}\arctan(x) - \frac{\pi}{4}\right)$. 

\begin{figure}[t!]
  \centering
  \includegraphics[width=\linewidth]{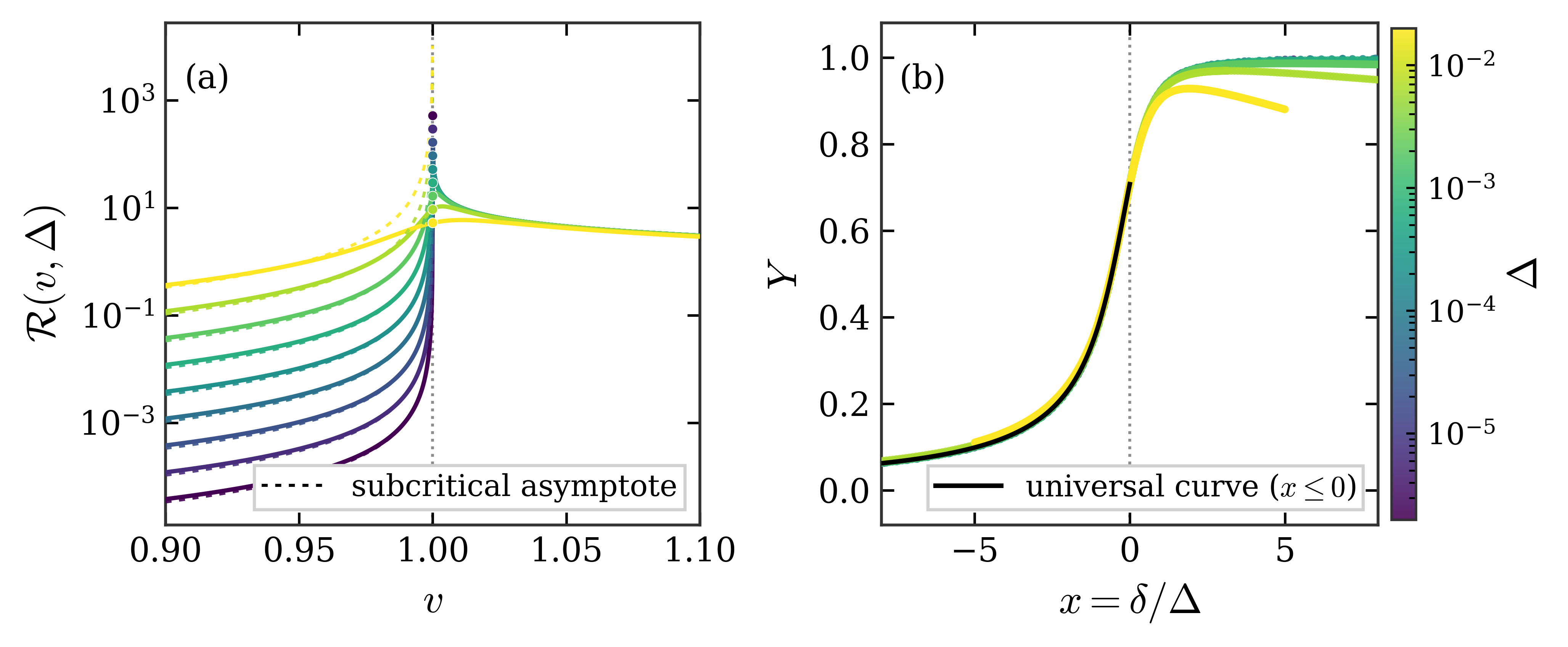}
\caption{\textbf{Threshold behavior near $v=1$.}
\textit{Left:} Mean resistance $\mathcal{R}(v,\Delta)$ in the threshold region for nine diffusivities, showing the peak at $v=1$ and Regime~III  asymptotics. \textit{Right:} Same data plotted against the reduced  variable $x=\delta/\Delta$ using the rescaled observable $Y=(2/\pi)\,e^{2a}\,\sqrt{S}\,\mathcal{R}$ with $S=\sqrt{\Delta^{2}+\delta^{2}}$,  as derived in Eq.~\eqref{eq:threshold}. All curves collapse onto the threshold profile for $x<0$, with supercritical deviations appearing as the two Havelock roots detach from $k=1$. Lorentzian pressure field with $a=0.20$.}
\label{fig:threshold}
\end{figure}

A broken symmetry between super and subcritical approaches is visible in Figs.~\ref{fig:R_vs_v} and \ref{fig:threshold} as sharp peaks near $v=1$ that broaden and diminish as diffusivity increases. The super/subcritical asymmetry reflects the underlying kernel structure. For $\delta > 0$, the two roots of $\omega(k) = vk$ detach from $k=1$. At high diffusivity, this threshold structure crosses over to the universal $\Delta^{-5/3}$ tail.

\textit{Discussion---}Stochastic fluctuations reshape the wave drag experienced at a fluid interface. In the deterministic theory, a source in uniform motion radiates and experiences resistance only when its speed exceeds $c_\text{min}$. Chepelianskii \emph{et al.}~\cite{chepelianskii2008capillary,closa2010capillary} previously showed that subcritical radiation also occurs for accelerated trajectories. Adding noise to the trajectory decorrelates the phase of wave emission and skews the mean surface profile around the source, and we have shown that this asymmetry is responsible for a net mean drag. Three quantitative predictions have been established, a universal high-diffusivity tail $\mathcal{R}\sim v\,\Delta^{-5/3}$ with geometry-independent prefactor, a threshold peak $\mathcal{R}(c_{\min},\Delta)\propto\Delta^{-1/2}$, and a subcritical law $\mathcal{R}\propto\Delta\,|\delta|^{-3/2}$, with the near-threshold response on the subcritical side collapsing onto the single variable $x=\delta/\Delta$ (Fig.~\ref{fig:threshold}).

The crossover from low to high diffusivity, occurring at $\Delta\sim1$, corresponds in water to $D\sim c_{\min}/k_c\sim6\times10^{-4}\,\mathrm{m^2\,s^{-1}}$. Thermal colloids lie at very small $\Delta$, whereas microscopic active tracers can access finite low-diffusivity values $\Delta<1$, making the subcritical and threshold regimes experimentally relevant. Values at or above the high-diffusivity crossover require strong active forcing, externally imposed trajectory noise, or macroscopic surface locomotion.

The framework developed herein applies to any stochastic trajectory whose stationary increments are known or can be estimated. When a closed form for the expectation in Eq.~\eqref{eq:gauss} is not available it can be evaluated from simulations or numerically provided the characteristic function $\varphi(k,\tau)$ is known. Thus, the formalism extends beyond analytically tractable cases. As a simple non-Gaussian example admitting a closed-form treatment, consider the drifted symmetric $\alpha$-stable L\'evy process, $x_t=vt+L_t^{(\alpha)}$, for which $\varphi(k,\tau)=e^{ikv\tau-D_\alpha \vert k\vert^\alpha \tau}$~\cite{klafter2011first}. Substitution into Eq.~\eqref{eq:kernel} reduces, after the $\tau-$integration, to the replacement $Dk^2\to D_\alpha k^\alpha$ in Eq.~\eqref{eq:kernel_BM_correct}, yielding the mean wave resistance in closed form given by Eq.~\eqref{eq:Rmean} with the modified kernel $\mathcal{K}(k;v,D) \to\mathcal{K}_\alpha(k;v,D_\alpha)=\frac{2 D_\alpha\, v\, k^{\alpha+1}}{((D_\alpha k^\alpha)^2 + (\omega-vk)^2)((D_\alpha k^\alpha)^2 + (\omega+vk)^2)}.$ The drifted Brownian result is recovered for $\alpha = 2$. For $\alpha<2$ the trajectory has discontinuous jumps at which the pressure field switches instantaneously position. The kernel captures these jumps directly through its dependence on the increment $x_t-x_{t-\tau}$. This extends wave-drag theory at fluid interfaces to active-matter trajectories and to surface foragers performing L\'evy flights~\cite{viswanathan1999optimizing,benichou2011intermittent}. 

\textit{Acknowledgments---}This project has received funding from the European Research Council (ERC) under the European Union’s Horizon 2020 research and innovation program (grant agreement No. 947731).

\bibliographystyle{apsrev4-2}
\bibliography{refs}

\end{document}


\title{Supplemental Material for\\
\emph{Wave Resistance for Stochastic Motion at Interfaces}}

\author{Maxence Arutkin}
\affiliation{School of Chemistry, Center for the Physics \& Chemistry of Living Systems, Sackler Center for Computational Molecular \& Materials Science, Tel Aviv University, 6997801, Tel Aviv, Israel}
\author{Shlomi Reuveni}
\affiliation{School of Chemistry, Center for the Physics \& Chemistry of Living Systems, Sackler Center for Computational Molecular \& Materials Science, Tel Aviv University, 6997801, Tel Aviv, Israel}
\author{Elie Rapha\"el}
\affiliation{Gulliver UMR CNRS 7083, ESPCI Paris, PSL Research University, 75005 Paris, France}

\date{\today}

\maketitle
\tableofcontents
\section{Linear capillary--gravity model and causal memory formula}

We consider a two-dimensional line source translated by a prescribed horizontal trajectory $x_t$ along the free surface $z=0$ of a deep, inviscid, incompressible, irrotational fluid. The vertical coordinate is oriented upward, the fluid occupies $z<0$, when $\eta(x,t)>0$ it denotes an upward surface displacement, and $P(x,t)>0$ denotes a downward external pressure applied by the source. With velocity field $\mathbf{u}=\nabla\phi$, the linearized boundary conditions at $z=0$ are
\begin{equation}
\partial_t \eta = \partial_z \phi\big|_{z=0}, \qquad
\partial_t \phi\big|_{z=0} + g\,\eta - \frac{\gamma}{\rho}\,\partial_x^2 \eta = -\,\frac{P(x,t)}{\rho}.
\label{eqS:BCs}
\end{equation}
A fixed pressure profile is translated along the prescribed trajectory,
\begin{equation}
P(x,t) = P_0\!\big(x - x_t\big), \qquad \widehat P(k,t) = \widehat P(k)\,e^{ik x_t},
\label{eqS:forcing}
\end{equation}
with Fourier convention $f(x)=\int_{\mathbb{R}} \frac{dk}{2\pi}\,\widehat f(k)\,e^{-ikx}$. Incompressibility together with irrotational flow implies that the velocity potential is harmonic in the bulk, $\Delta\phi=0$. Fourier modes then satisfy $(\partial_{z}^{2}-k^{2})\,\widehat\phi=0$. Deep-water boundedness as $z\to -\infty$ retains only the decaying branch of the solution, giving $\widehat\phi(k,z,t)=\varphi(k,t)\,e^{|k|z}$.
Eliminating $\varphi$ yields the forced oscillator
\begin{equation}
\ddot{\widehat \eta}(k,t) + \omega(k)^2\,\widehat \eta(k,t) = -\,\frac{|k|}{\rho}\,\widehat P(k)\,e^{ik x_t}, \qquad
\omega(k)^2 = g|k| + \frac{\gamma}{\rho}|k|^3.
\label{eqS:eta-oscillator}
\end{equation}
Causality selects the retarded solution
\begin{equation}
\widehat \eta(k,t) = -\,\frac{|k|\,\widehat P(k)}{\rho\,\omega(k)} 
\int_{0}^{t} \sin\!\left(\omega(k)(t-\tau)\right)\,e^{ik x_{\tau}}\,d\tau.
\label{eqS:eta-ret}
\end{equation}
With our convention, the wave resistance is the force that must be supplied to maintain the prescribed horizontal motion of the source. It is obtained by virtual work~\cite{havelock1932theory}. Under an infinitesimal horizontal translation $\delta x_t$ of the pressure distribution, the local elevation of the surface changes by $\partial_x\eta(x,t)\delta x_t$. The pressure force does the total virtual work
\begin{equation}
\delta W =-\delta x_t \!\int_{\mathbb{R}} P(x,t)\,\partial_x \eta(x,t)\,dx.
\end{equation}
Therefore, since the wave resistance is the force supplied opposite to this work
\begin{equation}
    \delta W = -R(t)\delta x_t
\end{equation}
hence,
\begin{equation}
R(t) = \!\int_{\mathbb{R}} P(x,t)\,\partial_x \eta(x,t)\,dx.
\label{eqS:R-def}
\end{equation}
Substituting Eq. (\ref{eqS:eta-ret}), carrying out the $x$-integral, and folding $k$ onto $k>0$ yields
\begin{equation}
R(t) = \frac{1}{\pi\rho}\int_{0}^{\infty}\!dk\,k^2\,\big|\widehat P(k)\big|^2
\int_{0}^{t}\!d\tau\,\frac{\sin\!\big(\omega(k)\tau\big)}{\omega(k)}\,
\sin\!\Big(k\,(x_t-x_{t-\tau})\Big).
\label{eqS:R-memory}
\end{equation}
For a sufficiently localized smooth pressure field the $k$-integral is absolutely convergent. The kernel is real and odd in the trajectory increment $x_t-x_{t-\tau}$.

\section{Ballistic limit and Havelock-type steady formula}

For uniform translation $x_t=vt$, define
\begin{equation}
I(k,v;t) = \int_{0}^{t}\!d\tau\, \frac{\sin(\omega\tau)}{\omega}\,\sin(kv\tau).
\label{eqS:I-def}
\end{equation}
In the long-time limit $t\to\infty$ one uses $\sin(\omega\tau)\sin(kv\tau)=\tfrac12[\cos((\omega-kv)\tau)-\cos((\omega+kv)\tau)]$ and $\int_{0}^{\infty}\cos(c\tau)\,d\tau=\pi\delta(c)$ to obtain
\begin{equation}
I(k,v;\infty) = \frac{\pi}{2\,\omega(k)}\,\delta\!\big(\omega(k)-kv\big).
\label{eqS:I-limit}
\end{equation}
Substitution into \eqref{eqS:R-memory} yields
\begin{equation}
R_{\mathrm{steady}}(v) = \frac{1}{2\rho}\int_{0}^{\infty}\!dk\,
\frac{k^2\,\big|\widehat P(k)\big|^2}{\omega(k)}\,\delta\!\big(\omega(k)-kv\big),
\label{eqS:Havelock-delta}
\end{equation}
nonzero only for $v\ge c_{\min}=(4g\gamma/\rho)^{1/4}$. Evaluating the delta on the roots $k_i(v)$ of $\omega(k)=vk$ gives
\begin{equation}
R_{\mathrm{steady}}(v)
= \sum_{i=1}^{2}\frac{k_i^2\,\big|\widehat P(k_i)\big|^2}{\rho\,\big|\,g+3\gamma k_i^{2}/\rho - 2v^2 k_i\,\big|} ,
\label{eqS:Havelock-final}
\end{equation}
this recovers the classical Havelock expression~\cite{havelock1932theory}, in the capillary–gravity formulation introduced by Rapha\"el and de~Gennes~\cite{raphael1996capillary}.

\section{Drifted Brownian: mean wave resistance and reduced kernel}

For $x_t= v t + B_t$ with $\langle B_t\rangle=0$ and $\langle (B_t-B_{t-\tau})^2\rangle=2D\,\tau$, the characteristic function is $\langle e^{ik(B_t-B_{t-\tau})}\rangle = e^{-Dk^2\tau}$. Averaging \eqref{eqS:R-memory} and taking $t\to\infty$ yields
\begin{equation}
\langle R\rangle(v,D)=\frac{1}{\pi\rho}\int_{0}^{\infty}\!dk\,k^2\,\big|\widehat P(k)\big|^2
\int_{0}^{\infty}\!d\tau\,\frac{\sin\!\big(\omega(k)\tau\big)}{\omega(k)}\,
\sin(k v\tau)\,e^{-D k^2\tau}.
\label{eqS:Rmean-raw}
\end{equation}
Writing $\sin(\omega\tau)\sin(kv\tau)=\tfrac12[\cos((\omega-kv)\tau)-\cos((\omega+kv)\tau)]$ and using $\int_{0}^{\infty}e^{-\alpha\tau}\cos(\xi\tau)\,d\tau = \alpha/(\alpha^{2}+\xi^{2})$ yields

\begin{equation}
\langle R\rangle(v,D)=\frac{2 v D}{\pi\rho}\int_{0}^{\infty}\!dk\,
\frac{k^{5}\,\big|\widehat P(k)\big|^{2}}
{\left((Dk^{2})^{2}+(\omega- v k)^{2}\right)\left((Dk^{2})^{2}+(\omega+ v k)^{2}\right)}.
\label{eqS:Rmean-physical}
\end{equation}

We rescale by capillary--gravity characteristic scales and normalize the forcing profile. Table~\ref{tab:rescaling} summarizes the transformation.

\begin{table}[h]
\centering
\begin{tabular}{@{} l c l @{}}
\toprule
\textbf{Quantity} & \textbf{Rescaling} & \textbf{Physical meaning} \\
\midrule
\multicolumn{3}{@{}l}{\textit{Intrinsic scales}} \\[2pt]
$k_c$ & $\sqrt{\rho g/\gamma}$ & Capillary--gravity crossover wavenumber \\
$\omega_{c}$ & $\sqrt{2gk_c}$ & Capillary--gravity crossover frequency \\
$c_{\min}$ & $\omega_{c}/k_c = (4g\gamma/\rho)^{1/4}$ & Minimum phase velocity \\[4pt]
\multicolumn{3}{@{}l}{\textit{Dimensionless variables}} \\[2pt]
$k\to$ & $k/k_c$ & Wavenumber (units of $k_c$) \\
$v\to$ & $v/c_{\min}$ & Drift speed (units of $c_{\min}$) \\
$\omega\to$ & $\omega/\omega_{c} = \sqrt{(k+k^3)/2}$ & Dispersion relation \\
$\Delta$ & $Dk_c/c_{\text{min}}$ & Diffusion strength \\
$\Phi(k)$ & $\widehat{P}(k)/\widehat{P}(0)$ & Normalized forcing ($\Phi(0)=1$) \\[4pt]
\multicolumn{3}{@{}l}{\textit{Resistance normalization}} \\[2pt]
$R_{\mathrm{eff}}$ & $\frac{2k_c^3}{\pi\rho\omega_{c}^2}|\widehat{P}(0)|^2$& Characteristic resistance scale \\
$\mathcal{R}$ & $\langle R\rangle/R_{\mathrm{eff}}$ & Dimensionless mean resistance \\
\bottomrule
\end{tabular}
\caption{\textbf{Dimensionless rescaling for Brownian wave resistance.} Intrinsic scales are set by gravity $g$, surface tension $\gamma$, and fluid density $\rho$. The dimensionless parameter $\Delta$ quantifies diffusive decorrelation relative to wave propagation timescales; $\Delta \ll 1$ corresponds to nearly ballistic motion, while $\Delta \gg 1$ indicates strong diffusion.}
\label{tab:rescaling}
\end{table}

This transformation reduces Eq.~\eqref{eqS:Rmean-physical} to the dimensionless form
\begin{equation}
\mathcal R(v,\Delta)=v\,\Delta\int_{0}^{\infty}\!dk\,
\frac{k^{5}\,|\Phi(k)|^{2}}
{\left((\Delta k^{2})^{2}+(\omega- v k)^{2}\right)\left((\Delta k^{2})^{2}+(\omega+ v k)^{2}\right)}.
\label{eqS:Rmean-dimless}
\end{equation}
From Eq.~\eqref{eqS:Rmean-dimless} onward, we keep the same symbols $k$, $\omega$, and $v$ for the dimensionless variables defined in Table~\ref{tab:rescaling}, and replace the dimensional diffusivity $D$ by the dimensionless diffusivity $\Delta$. Limiting behaviors: as $\Delta\to 0$ with $v>1$ fixed, $\mathcal{R}$ recovers the Havelock resistance via $\Delta k^2[(\Delta k^2)^2 + (\omega \mp vk)^2]^{-1} \to \pi\delta(\omega \mp vk)$; as $v\to 0$ at fixed $\Delta$, the prefactor ensures $\mathcal{R} \sim v$ vanishes linearly, consistent with zero mean drift.

\section{Mean profile}
From \eqref{eqS:eta-ret}, the same average over Brownian increments and time integration as in Sec.~S.III, taken in the stationary limit, give the Fourier-space mean elevation in the co-moving frame
\begin{equation}
\lim_{t\to\infty}\big\langle \widehat{\eta}(k,t)\,e^{-ikx_t}\big\rangle = -\,k\,\Phi(k)\,\frac{\left(\omega^{2}(k)-v^{2}k^{2}+\Delta^{2}k^{4}\right) - 2i\Delta v k^{3}}{\left(\omega^{2}(k)-v^{2}k^{2}+\Delta^{2}k^{4}\right)^{2} + \left(2\Delta v k^{3}\right)^{2}}.
\label{eqS:mean-profile-k}
\end{equation}

Taking the real part of the inverse Fourier transform gives the mean elevation in real space, centered on the instantaneous source position $X=x-x_t$:
\begin{equation}
\langle\eta(X)\rangle =\lim_{t\to \infty}\left\langle \eta(x_t+X,t)\right\rangle=-\frac{1}{\pi}\int_{0}^{\infty}\!dk\,\frac{k \, \Phi(k)\left(\left(\omega^2(k)-v^{2}k^{2}+\Delta^{2}k^{4}\right)\cos(kX)-2\Delta v k^{3}\sin(kX)\right)}{\left(\omega^2(k)-v^{2}k^{2}+\Delta^{2}k^{4}\right)^{2}+\left(2\Delta v k^{3}\right)^{2}}.
\label{eqS:mean-profile}
\end{equation}

\begin{figure}[h!]
\centering
\includegraphics[width=\linewidth]{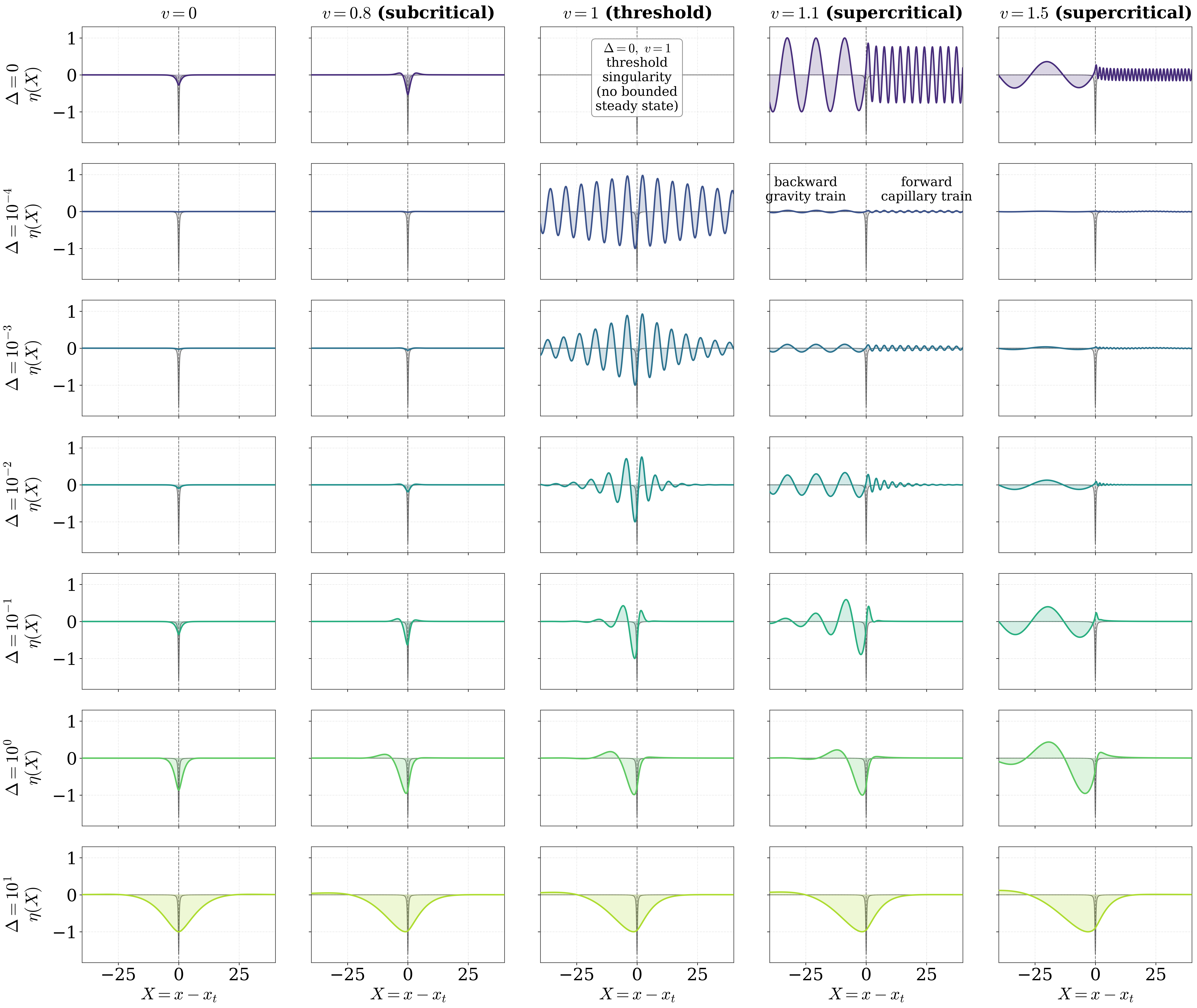}
\caption{Stationary mean elevation $\langle\eta(X)\rangle$ in the co-moving frame, from Eq.~\eqref{eqS:mean-profile} with $\Phi(k)=e^{-ak}$ and $a=0.2$. Columns: $v\in\{0,\,0.8,\,1,\,1.1,\,1.5\}$. Rows: $\Delta\in\{0,\,10^{-4},\,10^{-3},\,10^{-2},\,10^{-1},\,1,\,10\}$. Each row is rescaled to its row maximum. The gray region shows the pressure profile; the dashed line marks the source position $X=0$. The $(\Delta=0,\,v=1)$ panel has no bounded stationary solution.}
\label{figS:profile}
\end{figure}

Figure~\ref{figS:profile} shows, on a $(v,\Delta)$ grid, the stationary profile obtained by re-centering each realization on the instantaneous source position and then averaging. The drag is determined by the same averaged profile, Eq.~\eqref{eqS:R-def} yields
\begin{equation}
\langle R\rangle = \int_{\mathbb{R}} P_0(X)\,\partial_X\langle\eta(X)\rangle\,dX .
\end{equation}

When $\langle\eta(X)\rangle$  is symmetric around the source, the mean resistance integral vanishes if the pressure field is also symmetric. This is the case at $\Delta=0$ for $v<1$: the deformation is a symmetric dip and carries no mean drag. For $\Delta>0$, the average profile becomes skewed even below threshold, producing a finite mean resistance. For $v>1$ and $\Delta=0$, the deterministic capillary and gravity trains are recovered and generate the Havelock drag. Increasing $\Delta$ regularizes the threshold and damps the oscillatory profiles. The damping is asymmetric, both trains drift away from the source with the same absolute relative speed with respect to the source, but these trains decohere at different rates. The source's Brownian displacement during a given time interval represents a much larger fraction of the short capillary wavelength than of the longer gravity wavelength. The same fluctuation around the straight drift trajectory therefore scrambles the capillary phase much faster. The capillary waves run out of coherence over a shorter distance than the gravity waves, which is why the figure shows that the capillary waves are suppressed at smaller $\Delta$ than the gravity waves. At large $\Delta$, both trains are gone and the profile becomes an asymmetric dip for $v>0$.

\section{Derivation of Eq. (5)}

From \eqref{eqS:Rmean-dimless}, in the contributing small-$k$ sector (large $\Delta$) take $\omega(k)^{2}=k/2$ and $|\Phi(k)|^{2}\to 1$. With $k=\Delta^{-2/3}y$,
\begin{equation}
\mathcal R(v,\Delta)\sim
v\,\Delta^{-5/3}\int_{0}^{\infty}\frac{y^{5}}{\big(y^{4}+y/2\big)^{2}}\,dy.
\label{eqS:RegI-inner}
\end{equation}
This integral evaluates to
\begin{equation}
\int_{0}^{\infty}\frac{y^{5}}{\big(y^{4}+y/2\big)^{2}}\,dy
= \frac{2^{5/3}\,\pi}{9\sqrt{3}}.
\label{eqS:RegI-const}
\end{equation}
Therefore
\begin{equation}
\mathcal R(v,\Delta)\sim C_\star\,v\,\Delta^{-5/3}, \qquad
C_\star=\frac{2^{5/3}\pi}{9\sqrt{3}}.
\label{eqS:RegI-result}
\end{equation}

\section{Derivation of Eq. (6) and asymptotic analysis}

With $|\Phi(k)|^{2}=e^{-2 a k}$ ($a>0$), at leading order in $v,\Delta$ (set $v=0$ in denominators but keep $\Delta$ to ensure convergence at large $k$),
\begin{equation}
\mathcal R(v,\Delta)\sim v\,\Delta\,A_{0}(a,\Delta), \qquad
A_{0}(a,\Delta)=\int_{0}^{\infty}\frac{k^{5}\,e^{-2 a k}}{\big((\Delta k^{2})^{2}+\omega(k)^{2}\big)^{2}}\,dk.
\label{eqS:RegII-leading}
\end{equation}
Using $\omega(k)^{2}=\tfrac12 k(1+k^{2})$ gives
\begin{equation}
A_{0}(a,\Delta)=4\int_{0}^{\infty}
\frac{k^{3}\,e^{-2 a k}}{\big(1+k^{2}+2\Delta^{2}k^{3}\big)^{2}}\,dk.
\label{eqS:RegII-exact}
\end{equation}
Asymptotics:

(i) $a\to0$ with $a\gg\Delta^{2}$,
\begin{equation}
A_{0}(a,\Delta)\sim4\log\!\frac{1}{a}-4(\gamma_{E}+\log 2)-2.
\label{eqS:RegII-case1}
\end{equation}

(ii) $a\to0$ with $a\ll\Delta^{2}$,
\begin{equation}
A_{0}(a,\Delta)\sim 8\log\!\frac{1}{\Delta}-4\log 2 - 6.
\label{eqS:RegII-case2}
\end{equation}

(iii) $a\to\infty$ (small-k dominated regime)
\begin{equation}
\mathcal R(v,\Delta)\sim \frac{3}{2}\,v\,\Delta\,a^{-4}.
\label{eqS:RegII-case3}
\end{equation}
In the logarithmic cases the observable $v\,\Delta\,A_{0}$ vanishes as $\Delta\to0$.

\section{Derivation of Eq. (7)}

This section treats the threshold point $v=1$ and its subcritical side. We write $\delta=v-1\le0$ with $|\delta|\ll1$. The supercritical case $v>1$ is not expanded here. For $v<1$ there are no Havelock roots. At $v=1$, the two ballistic roots coalesce at the degenerate threshold point $k=1$, and the finite-$\Delta$ kernel regularizes the corresponding singularity.

Start from
\begin{equation}
\mathcal R(v,\Delta)=\frac{1}{2}\int_{0}^{\infty}\!d\tau\int_{0}^{\infty}\!dk\;
\frac{k^{2}\,e^{-2ak}}{\omega(k)}\,e^{-\Delta k^{2}\tau}\,
\sin(kv\tau)\,\sin(\omega\tau), \qquad \omega(k)=\sqrt{\frac{k+k^{3}}{2}}.
\label{eqS:RegIII-start}
\end{equation}
With $\sin A\,\sin B=\tfrac12[\cos(A-B)-\cos(A+B)]$ define the phase functions $\varphi_{\pm}(k)=\omega(k)\pm kv $ and $F(k)=k^{2}e^{-2ak}/\omega(k)$. For fixed $\tau$, $\varphi_{-}$ has a stationary point at $k_{\text{sp}}$ near $k=1$ defined by $\omega'(k_{\text{sp}})=v$; the phase $\varphi_{+}$ has no stationary point for $k>0$ in this region. For $v=1+\delta$ with $\delta\le 0$ and $|\delta|\ll 1$, $k_{\text{sp}}$ lies near $1$ and admits the local expansion
\begin{equation} 
k_{\text{sp}}=1+2\delta+2\delta^{3},\qquad
\Psi_{\text{sp}}=k_{\text{sp}} v-\omega(k_{\text{sp}})=\delta+\delta^{2},\qquad
\omega''(k_{\text{sp}})=\tfrac12-\tfrac{3}{2}\delta^{2}.
\end{equation}
A Gaussian evaluation at fixed $\tau$ gives
\begin{equation}
I(\tau)\sim F(k_{\text{sp}})\,e^{-\Delta k_{\text{sp}}^{2}\tau}\,
\sqrt{\frac{2\pi}{\omega''(k_{\text{sp}})\,\tau}}\,
\cos\!\Big(\tau\Psi_{\text{sp}}-\frac{\pi}{4}\Big).
\label{eqS:RegIII-I}
\end{equation}
Using $\int_{0}^{\infty}\tau^{-1/2}e^{-\alpha\tau}\cos(\beta\tau-\tfrac{\pi}{4})\,d\tau
=\sqrt{\pi}\,(\alpha^{2}+\beta^{2})^{-1/4}\,
\cos\!\big(\tfrac12\arctan(\beta/\alpha)-\tfrac{\pi}{4}\big)$ yields
\begin{equation}
\mathcal R(v,\Delta)\sim
\frac{\pi}{2\sqrt2}\,\frac{F(k_{\text{sp}})}{\sqrt{\omega''(k_{\text{sp}})}}\,
\frac{\cos\!\Big(\tfrac12\arctan\!\dfrac{\Psi_{\text{sp}}}{\Delta k_{\text{sp}}^{2}}-\tfrac{\pi}{4}\Big)}
{\left[(\Delta k_{\text{sp}}^{2})^{2}+\Psi_{\text{sp}}^{2}\right]^{1/4}}.
\label{eqS:RegIII-uniform}
\end{equation}
Near threshold one may take $k_{\text{sp}}\to1$, $F(k_{\text{sp}})\to e^{-2a}$, $\omega''(k_{\text{sp}})\to\tfrac12$, $\Psi_{\text{sp}}\to\delta$, giving
\begin{equation}
\mathcal R(v,\Delta)\sim \frac{\pi}{2}\,e^{-2a}\,
\frac{\cos\!\Big(\tfrac12\arctan\!\dfrac{\delta}{\Delta}-\tfrac{\pi}{4}\Big)}
{(\Delta^{2}+\delta^{2})^{1/4}}.
\label{eqS:RegIII-simple}
\end{equation}
With $\theta=\arctan(\delta/\Delta)$ and $S=\sqrt{\Delta^{2}+\delta^{2}}$, this becomes
\begin{equation}
\mathcal R(v,\Delta)\sim\frac{\pi}{4}\,e^{-2a}\,
\frac{\sqrt{S+\Delta}-\,\sqrt{S-\Delta}}{S}, \qquad S=\sqrt{\Delta^{2}+\delta^{2}}.
\label{eqS:RegIII-radical}
\end{equation}

At threshold,
\begin{equation}
\mathcal R(1,\Delta)\sim
\frac{\pi}{2\sqrt2}\,e^{-2a}\,\Delta^{-1/2},
\qquad \Delta\ll1.
\end{equation}
For $\delta<0$, with $\Delta\ll|\delta|\ll1$,
\begin{equation}
\mathcal R(v,\Delta)\sim
\frac{\pi}{4}\,e^{-2a}\,\Delta\,|\delta|^{-3/2}.
\end{equation}

\section{Summary table of asymptotic regimes}

\begin{table}[h]
\centering
\footnotesize
\setlength{\tabcolsep}{4.5pt}
\renewcommand{\arraystretch}{1.15}
\begin{tabular}{@{} l l l l @{}}
\toprule
\textbf{Regime} & \textbf{Asymptotic conditions} & \textbf{Leading asymptotic for $\mathcal R(v,\Delta)$} & \textbf{Scaling exponents} \\
\midrule
I & $v$ fixed,\ $\Delta\to\infty$ &
$\mathcal R \sim C_\star\, v\, \Delta^{-5/3},\quad C_\star=\dfrac{2^{5/3}\pi}{9\sqrt{3}}$ &
$v^{1}\,\Delta^{-5/3}$ \\
\midrule
II(i) & $v\to0$, $\Delta\to0$;\ $a\to0$, $a\gg \Delta^{2}$ &
$\mathcal R \sim v\,\Delta\!\left(\,4\ln(1/a)-4(\gamma_E+\ln2)-2\,\right)$ &
$v^{1}\,\Delta^{1}$ \\
II(ii) & $v\to0$, $\Delta\to0$;\ $a\to0$, $a\ll \Delta^{2}$ &
$\mathcal R \sim v\,\Delta\!\left(\,8\ln(1/\Delta)-4\ln2-6\,\right)$ &
$v^{1}\,\Delta^{1}\log(\Delta^{-1})$\\
II(iii) & $v\to0$, $\Delta\to0$;\ $a\to\infty$ &
$\mathcal R \sim \tfrac{3}{2}\, v\,\Delta\, a^{-4}$ &
$v^{1}\,\Delta^{1}$\\
\midrule
III(a) & $v=1$,\ $\Delta\ll1$ &
$\mathcal R \sim \dfrac{\pi}{2\sqrt{2}}\,e^{-2a}\,\Delta^{-1/2}$ &
$(\delta,\Delta)$: $\Delta^{-1/2}$ at $\delta=0$ \\
III(b) & $v=1+\delta$, $\delta<0$, $\Delta \ll|\delta|\ll 1$&
$\mathcal R \sim \dfrac{\pi}{4}\,e^{-2a}\,\Delta\,|\delta|^{-3/2}$ &
$(\delta,\Delta)$: $\Delta^{1}\,|\delta|^{-3/2}$ \\
\bottomrule
\end{tabular}
\caption{Asymptotic regimes for $\mathcal R(v,\Delta)$ with $|\Phi(k)|^2=e^{-2ak}$. Exponents are given in $v,\Delta$ (and $\delta=v-1$ where relevant).}
\label{tabS:regimes}
\end{table}

\bibliographystyle{apsrev4-2}
\bibliography{refs}